\documentclass[12pt]{article}
\usepackage{fullpage,cmu-titlepage2}
\usepackage{times}

\ifx\pdftexversion\undefined
  \usepackage[dvips]{graphics}
\else
  \usepackage[pdftex]{graphics}
\fi

\usepackage{amsmath}
\usepackage{rotating}
\usepackage{tabularx}
\usepackage[
pageanchor=true,
plainpages=false,
pdfpagelabels, 
bookmarks,
bookmarksnumbered,
pdfborder=0 0 0,
pdfpagemode=UseOutlines]{hyperref}
\usepackage{draftstamp}
\usepackage{subfigure}

\title{Compactness via Pattern Stepping Bisimulation}

\author{Matias Scharager}

\date{April 2024}

\abstract{
The compactness lemma in programming language theory states that any recursive function can be simulated by a finite unrolling of the function. One important use case it has is in the {\em logical relations} proof technique for proving properties of typed programs, such as strong normalization. The relation between recursive functions and their finite counterparts is a special variant of the class of bisimulation relations. However, standard bisimulation proof approaches do not apply to the compactness lemma as properties of the relation vary over execution. As a result, the proof of compactness is often messy because the multiple copies made of the recursive function during execution can be unrolled an inconsistent number of times.
We present a new proof technique by indexing the bisimulation relation over the step transitions and utilizing an intermediate ``pattern'' language to mechanize bookkeeping. This generalization of ``pattern stepping bisimulation'' obviates the need for contextual approximation within the compactness lemma, and thus extends the compactness lemma to a wider range of programming languages, including those that incorporate control flow effects. We demonstrate this approach by formally verifying the compactness lemma within the Coq theorem prover in the setting of explicit control flow and polymorphism.
}


\trnumber{CMU-CS-24-117}

\citationinfo{
}

\usepackage{mathpartir}
\usepackage{xcolor}
\usepackage[frozencache=true,cachedir=minted-cache]{minted}
\usepackage{cleveref}
\usepackage{amsthm}

\makeatletter

\DeclareDocumentCommand \inferrule { s O {} m m o }{%
  \IfBooleanTF{#1}%
  {%
    \mpr@inferstar[#2]{#3}{#4}%
  }{%
    \mpr@inferrule[#2]{#3}{#4}%
  }%
  \IfValueT{#5}%
  {%
    \ 
    #5%
    \my@name@inferrule{#5}%
  }%
}
\NewDocumentCommand \my@name@inferrule { m }{%
  \def\@currentlabelname{\ensuremath{#1}}%
}
\makeatother

\usepackage{color}

\newcommand{\fhole}[0]{\omega}
\newcommand{\iin}{\ \mathtt{in}\ }
\newcommand{\llet}{\mathtt{let}\ }
\newcommand{\as}{\ \mathtt{as}\ }
\newcommand{\pack}[0]{\mathtt{pack}\ }
\newcommand{\open}[0]{\mathtt{open}\ }

\newcommand{\letcc}[0]{\mathtt{letcc}}
\newcommand{\abort}[0]{\mathtt{abort}}
\newcommand{\throw}[0]{\mathtt{throw}}
\newcommand{\cont}[0]{\mathtt{cont}}
\newcommand{\val}{\mathtt{val}}
\newcommand{\valf}{\val^\fhole}
\newcommand{\is}{\ \mathtt{is}\ }
\newcommand{\steps}{\longmapsto}

\newcommand{\stepsf}{\steps_\fhole}
\newcommand{\terminates}[0]{\downarrow}

\newcommand{\hole}{[\cdot]}
\newcommand{\fun}[0]{\mathtt{fun}\ }

\newcommand{\Tfun}[2]{#1\to#2}

\newcommand{\Texist}[2]{\exists #1. #2}
\newcommand{\Tcont}[1]{\cont(#1)}

\newcommand{\tmfun}[5]{\fun #1(#2:#3):#4 \is #5}
\newcommand{\tmapp}[2]{(#1)(#2)}
\newcommand{\tmLam}[2]{\Lambda #1. #2}
\newcommand{\tmApp}[2]{(#1)[#2]}
\newcommand{\tmpack}[4]{\pack (#1, #2) \as \Texist{#3}{#4}}
\newcommand{\tmopen}[4]{\open (#1) \as (#2,#3) \iin (#4)}
\newcommand{\tmpair}[2]{\langle #1, #2\rangle}
\newcommand{\tmprojl}[1]{\pi_1(#1)}
\newcommand{\tmprojr}[1]{\pi_2(#1)}
\newcommand{\tmlet}[3]{\llet #1 = (#2) \iin #3}
\newcommand{\tmletcc}[3]{\letcc\{#1\}(#2.#3)}
\newcommand{\tmthrow}[2]{\throw(#1;#2)}
\newcommand{\tmabort}[2]{\abort\{#1\}(#2)}
\newcommand{\tmcont}[1]{\cont(#1)}

\newcommand{\subst}[3]{[#1/#2]#3}
\newcommand{\substpair}[5]{[#1,#3/#2,#4]#5}

\newcommand{\static}[3]{#1\vdash #2 : #3}
\newcommand{\ESJ}[3]{#1 : (\cdot\vdash #2)\Rightarrow (\cdot\vdash #3)}

\newcommand{\of}[3]{#2\ \mathbf{of}^{#1}\ #3}

\definecolor{toPat}{rgb}{1,0,0}
\definecolor{toStep}{rgb}{0,0,1}

\newtheorem{theorem}{Theorem}
\newtheorem{lemma}[theorem]{Lemma}
\newtheorem{corollary}[theorem]{Corollary}

\begin{document}
\renewcommand*{\thepage}{title-\arabic{page}} 
\maketitle
\renewcommand*{\thepage}{\arabic{page}} 

\section{Introduction}\label{section:intro}

The name ``compactness'' originates from domain theoretic topics dealing with infinite objects.
When we encode a function in a program, we define an output for all possible inputs to the function, even if the collection of all inputs is infinite.
The key idea of compactness is that we are only capable of observing a finite amount of this information during program execution that eventually terminates.

The compactness lemma of this paper is this pattern of reasoning applied to the recursive depth of a function. This helps bring several key properties of denotational semantics and domain theory into the realm of operational semantics, bridging the gap between the representations of programming languages \cite{mason1996operational}.

The central principle behind compactness is how the recursive depth of the function changes during execution. Within a terminating program, the function had to have recursively called itself up to a finite recursive depth, so replacing the function with a finite unrolling with the same behavior up to that recursive depth should result in the same outcome.

While the compactness lemma is easy to believe informally, the need for excessive bookkeeping on the depth of all the instances of the recursive function during execution makes it difficult to break down the lemma into easily verified components.

\subsection{Compactness as a Bisimulation}

Consider a state transition system $(S,\to)$ where $\to \subseteq S\times S$ represents the transition function between states.
A bisimulation relation $R$ is a binary relation $R\subseteq S\times S$ where for any related states $p\ R\ q$ the following two properties hold:
\begin{enumerate}
    \item $p \to p'$ implies there exists $q\to q'$ such that $p'\ R\ q'$
    \item $q \to q'$ implies there exists $p\to p'$ such that $p'\ R\ q'$
\end{enumerate}
Transitioning one side of the bisimulation relation ensures that we can find a transition for the other side that maintains the symmetric relation.

This strategy extends cleanly to the operational analysis of programming languages as we can express our small-step semantics as the transition function within bisimulation \cite{sangiorgi2011introduction}.
This allows us to relate two programs $e$ and $d$ by showing that they are related $e\ R\ d$ after each single step of execution.
This is particularly useful for proving co-termination as if one program terminates, we can show that the other program must also terminate by inducting on the stepping relation.

This presentation of bisimulation doesn't account for the case of the compactness lemma.
We can define a relation that relates a recursive function to its unrolling of depth $n$, however, this is not a standard bisimulation relation because the unrolling depth decreases through execution whenever the function is called.

Instead, it is reasonable to consider a chain of binary relations $R_1\subseteq R_2 \subseteq \cdots$ where the following two properties hold for $p\ R_n\ q$:
\begin{enumerate}
    \item $p \to p'$ implies there exists $q\to q'$ such that $p'\ R_{n+1}\ q'$
    \item $q \to q'$ implies there exists $p\to p'$ such that $p'\ R_{n+1}\ q'$
\end{enumerate}
In this setting, our relation $R_n$ is gradually getting weaker over time as we execute, but it otherwise follows the same pattern of bisimulation.

We find that compactness more closely matches this type of bisimulation, and we will later substantiate this claim with the $\mathbf{of}^{n}$ relation in \cref{section:of}.

\subsection{Practical Application of Compactness}

In practice, the compactness lemma is utilized primarily for proving important properties of logical relations.
Logical relations are equivalences built inductively on type structure that allows for reasoning about various behavioral properties of programs.
The theorem that enables the use of these logical relations is known as the fundamental theorem of logical relations (FTLR) which states that all well-typed terms belong to the logical relation.
Recursive functions create circular reasoning within the proof of FTLR, and compactness resolves this circular reasoning by allowing us to induct on the finite recursive depth of the function.

A detailed example of how compactness (referred to as the unwinding theorem) is used to verify FTLR can be found in the seventh chapter of the Advanced Topics in Types and Programming Languages textbook \cite{pitts2005typed}.
We will briefly summarize this approach to showcase the usefulness of compactness.

Suppose that we have some logical relation $R$ over closed values of the language defined recursively on the types where the case for recursive functions appears as some variation of 
\begin{align*}
(\tmfun{&f}{x}{A}{B}{e})\in R[\Tfun{A}{B}]=\\
&\forall v\in R[A].\ (\substpair{\tmfun{f}{x}{A}{B}{e}}{f}{v}{x}{e})\in R[B].
\end{align*}

To prove that $F\in R[\Tfun{A}{B}]$, we need to show that a substitution of $[F/f]$ is a related substitution under $R$, which requires showing that $F\in R[\Tfun{A}{B}]$, hence creating a circular argument.

Now suppose that $F_n$ is the $n$th unwinding of the recursive function where it exactly matches the behavior of $F$ up to a recursive depth of $n$.
We can break free of the circular argument by inducting on this unwinding through the admissibility theorem.
\begin{theorem}[Admissibility]
If for all $n$, $F_n\in R[\Tfun{A}{B}]$, then we have $F\in R[\Tfun{A}{B}]$.
\end{theorem}

Finally, to prove admissibility, we need to show the compactness lemma: any instance of $F$ within a whole program can be computationally modeled by a specific unrolling $F_i$.
Since admissibility holds for all $n$, then the specific $i$ we care about is covered by the assumption, and admissibility holds.

\section{Formally Expressing Compactness}

\begin{figure}
    \centering
    \begin{align*}
        \tau &:= \cdots \mid \tau \to \tau \\
        e &:= \cdots \mid \tmfun{f}{x}{\tau}{\tau}{e}\mid e\ e
    \end{align*}
    \begin{mathpar}
    \inferrule{
        \static{\Gamma, f:\Tfun{\tau_1}{\tau_2}, x : \tau_1}{e}{\tau_2}
    }{
        \static{\Gamma}{\tmfun{f}{x}{\tau_1}{\tau_2}{e}}{\Tfun{\tau_1}{\tau_2}}
    }[\Tfun{}{}I]\label{tsj:fun}\and
    \inferrule{
        \static{\Gamma}{e_1}{\Tfun{\tau_2}{\tau}}\\
        \static{\Gamma}{e_2}{\tau_2}
    }{
        \static{\Gamma}{e_1\ e_2}{\tau}
    }[\Tfun{}{}E]\label{tsj:app}\and
    \inferrule{
        \quad
    }{
        \tmfun{f}{x}{\tau}{\tau'}{e'}\ \val
    }[\fun\val]\label{val:fun}\and
    \inferrule{
        e\ \val
    }{
        \tmapp{\tmfun{f}{x}{\tau}{\tau'}{e'}}{e}\steps
        \substpair{\tmfun{f}{x}{\tau}{\tau'}{e'}}{f}{e}{x}{e'}
    }[\Tfun{}{}\steps]\label{dyn:fun}\and
    \end{mathpar}
    \caption{Relevant components of the programming language}
    \label{fig:language}
\end{figure}

To properly define the compactness lemma, we must first define a programming language with recursive functions to work with. The relevant syntax we use can be found in \cref{fig:language} and we utilize the standard call-by-value dynamics for this language. This proof strategy does not rely on any specific definition of a programming language, but it requires the following properties of the reduction semantics.

\begin{lemma}[Determinism]\label{lemma:determinism}\quad
\begin{enumerate}
    \item If $e\steps a$ and $e\steps b$ then $a=b$
    \item Not both of $e\ \val$ and $e\steps e'$
\end{enumerate}
\end{lemma}

\begin{lemma}[Safety]\label{lemma:safety} For all well-formed programs $e$, either there exists some well-formed program $e'$ such that $e\steps e'$ or $e\ \val$.\end{lemma}

The specific definition of ``well-formed'' we utilize is that of a typing judgment and it is standard practice to prove the type safety of typed languages.
However, this formalization of compactness does not directly rely on the types themselves and can be extended to an un-typed setting, as long as there is a meaningful property of ``well-formed'' that shows this is a reduction system.

Fix an arbitrary closed function $\tmfun{f}{x}{\tau_1}{\tau_2}{e_f}$ that we wish to represent by its unrolling via compactness. This parameter is passed along to the rest of the formalization as input arguments so that the formalization can abstract over the function in question.

We also create more convenient syntax for expressing multiple steps taken. $e\steps^n e'$ signifies that $e$ steps to $e'$ in $n$ number of steps for a natural number $n$. The judgment $e\terminates$ signifies that $e$ terminates, or more formally that there exists a program $v$ and natural number $n$ such that $e\steps^n v$ and $v\ \val$.

Much like prior works, we define function unrollings with the syntax of \cref{fig:unrolling}. $F_{\fhole}$ serves as a convenient notation for the full recursive function. $F_{n}$ represents an unrolling that matches the full recursive function up to a depth of $n$. The base case $F_0$, commonly referred to as $\lambda.\bot$, is the simplest way to express a function that will loop infinitely when called, allowing for an easy way of expressing non-termination. We are able to prove the non-termination of $F_0$ calls directly as a lemma.

\begin{figure}
    \centering
    \begin{align*}
        F_0 &= \tmfun{g}{y}{\tau_1}{\tau_2}{\tmapp{g}{y}}\\
        F_{n+1} &= \tmfun{\_}{y}{\tau_1}{\tau_2}{\substpair{F_n}{f}{y}{x}{e_f}}\\
        F_\omega &= \tmfun{f}{x}{\tau_1}{\tau_2}{e_f}
    \end{align*}
    \caption{Function unrollings}
    \label{fig:unrolling}
\end{figure}

\begin{lemma}[Bottom]\label{lemma:inf_loop} $\tmapp{F_0}{v}$ does not terminate for any $v\ \val$.
\end{lemma}

\Cref{lemma:inf_loop} is the only place in this formalization reliant on determinism, \cref{lemma:determinism}.
This means this method of proving compactness can extend to any non-deterministic language that admits \cref{lemma:inf_loop}.
The only modification needed is that the precise definition of termination must be altered either to ``there exists a path that terminates'' or ``all paths terminate,'' whichever one is desired by the language designer.

With all this in place, we can express our end goal of the compactness lemma for our chosen function. 

\begin{lemma}[Compactness]\label{lemma:compactness}$\subst{F_\omega}{x}{e}\terminates\ \mathrm{iff}\ \exists n.\ \subst{F_n}{x}{e}\terminates$.\end{lemma}

This is not the only definition of compactness that this formalization gives us, but it is strong enough to prove the admissibility of a logical relation which is what the compactness lemma is normally used for in practice.

\section{Pattern Language Definition}

Abstracting over the function we care about, we can now define a pattern language that extends our original language as defined in \cref{fig:DynExtension}.
All other dynamic rules are recursively defined in the same way, replacing all instances of $\val$ and $\steps$ with $\valf$ and $\stepsf$ respectively.

\begin{figure}
    \centering
    \begin{align*}
        \tau &:= \cdots \mid \tau \to \tau \\
        e &:= \cdots \mid \textbf{fun}\ f\ (x : \tau) : \tau.\ e\mid e\ e \mid \fhole
    \end{align*}
    \begin{mathpar}
    \inferrule{
        \quad
    }{
        \static{\Gamma}{\fhole}{\Tfun{\tau_1}{\tau_2}}
    }[\fhole I]\label{tsj:hole}\and
    \inferrule{
    }{
        \fhole\ \valf
    }[\fhole\ \valf]\label{val:hole}\and
    \inferrule{
        e\ \valf
    }{
        \tmapp{\fhole}{e}\stepsf \substpair{\fhole}{f}{e}{x}{e_f}
    }[\fhole\stepsf]\label{dyn:hole}\and
    \end{mathpar}
    \caption{Pattern stepping extension under $\tmfun{f}{x}{\tau_1}{\tau_2}{e_f}$}
    \label{fig:DynExtension}
\end{figure}

The $\fhole$ is a reserved variable name that is global in scope, and unlike other variables that occur locally in $\Gamma$, it has associated dynamics.
A ``closed'' program within this language can contain the free variable $\fhole$ because the dynamics define how to handle the execution of this variable.
We refer to programs in this language as ``patterns'' because they look like terms with $\fhole$ holes in them to pattern match a whole term.

This strategy is inspired by the idea of a program capsule \cite{jeannin2012computing} where closed programs feature the pair $\langle e, \sigma\rangle$ where $e$ is an open term and $\sigma$ is a substitution context which provides meaning to the variables in $e$.
In this setting, the substitution context $\sigma$ can be thought of as a ``stack'' of immutable memory, so the open term $e$ can continue executing normally by reading the values stored for each free variable on the stack.
This framework enables a nice separation between the computation components of a term $e$ and the value components located in the stack $\sigma$.

For our purposes, we only allow the one free variable $\fhole$, so our stack is constructed as $\fhole\hookrightarrow F_\fhole$.
Replacing $F_\fhole$ with $F_n$ for the compactness argument affects only the stack and not the execution of the open term $e$.
Hence, we can easily prove compactness by showing that the overall pattern of execution for open term $e$ remains unchanged under related definitions of the stack.

\subsection{Relating Terms to Patterns}\label{section:of}

It is easy to see a conversion between a capsule $\langle e,\sigma\rangle$ and a standard program by carrying out the substitution into the open term $\sigma(e)$.
In our formalization, we recreate this substitution explicitly to relate closed terms of our original language to patterns of the pattern language.

We define $\of{n}{e}{p}$ to be the compatible relation between terms and patterns admitting $\of{n}{F_{\omega}}{\fhole}$ and for all $i\geq n$, $\of{n}{F_{i}}{\omega}$.
As such, $\of{n}{e}{p}$ signifies that term $e$ exactly matches pattern $p$ except in the specific places that $\fhole$ is in $p$.
Moreover, these specific places in $e$ must have unrollings of the function of a depth of at least $n$, specifying the minimum unrolling of the term.
We choose the name $\mathbf{of}$ because we see that the term is ``of'' the pattern, in that the term pattern matches based on the marked locations in the pattern.

The benefit of this definition is that it condenses bookkeeping all instances of the unrolling into maintaining a singular number representing the minimum possible unrolling that we can use for induction in later proofs. There is no restriction that the minimum has to exist, but simply that all existing unrollings are greater or equal to this minimum.

Maintaining only the minimum presents an underestimation of the present function unrollings. Our proofs must always assume the worst case of the minimum function unrolling, thus often requiring a greater depth of recursion than is actually needed. However, the burden of this inefficiency affects only the mathematical formalization rather than the program execution itself, and the compactness lemma cares only about the existence of a depth of unrolling without caring about the minimum such depth, hence there's no issue.

This flexibility allows us to arbitrarily decrease the minimum counter whenever we deem it necessary, and we express this property explicitly via \cref{lemma:dec_of}. This creates an explicit ordering on the $\mathbf{of}^n$ relation based on the natural number $n$ where the weakest such relation is $\mathbf{of}^0$ where any level of function unrollings is allowed. Later proofs often give us the stronger inductive hypothesis of $\of{n+1}{e}{p}$ for some inductive cases when what we require is the weaker statement of $\of{n}{e}{p}$, so the following lemma is used for those cases.

\begin{lemma}[Decrement Pattern]\label{lemma:dec_of} $\of{n+1}{e}{p}$ implies $\of{n}{e}{p}$.
\end{lemma}

Because of this ordering, we can see that $\of{n}{e}{p}$ behaves as the described ``indexed bisimulation relation'' of \cref{section:intro}.
We start with the stronger relation $\mathbf{of}^{n+1}$ and then decrement to the weaker $\mathbf{of}^{n}$ as we execute.

\section{Proof Strategy}

The main principle of the proof strategy for compactness is an indexed bisimulation between term and pattern dynamic judgments expressed by the $\mathbf{of}^n$ relation. Under certain assumptions expressed in the lemmas, we can convert between $\val$ and $\valf$ and between $\steps$ and $\stepsf$. We can think of the forward direction as clearing out all instances of the function with holes in the pattern, and then the backward direction is filling in the holes with specific unrollings of the function we care about, all of this without affecting the rest of the structure of the judgments.

We start first by relating $\val$ and $\valf$.
This is necessary for the overall proof of compactness and also within the later lemmas if the definition of $\steps$ is dependent on $\val$ as in the call-by-value setting.
If we have $\of{n}{e}{p}$ and focus specifically on the places where $e$ and $p$ are distinct, we end up with $F_i$ on the left and $\fhole$ on the right, both of which are values by $\fun\ \val$ and $\fhole\ \valf$, making the two judgments match up.

\begin{lemma}[Value Pattern]\label{lemma:val_valf} Given $\of{n}{v}{v_p}$, then $v\ \val$ iff $v_p\ \valf$.
\begin{proof}
In the forward direction, solve by induction on $v\ \val$ and inversion on $\of{n}{v}{v_p}$. Similarly, in the backwards direction, solve by induction on $v_p\ \valf$ and inversion of $\of{n}{v}{v_p}$
\end{proof}
\end{lemma}

\begin{figure}
    \centering
    \begin{tabular}{clc}
        $e$ & $\steps$ & $e'$\\
        \hspace{.2em}$\of{0}{}{}$& &{\color{toPat} $\of{0}{}{}$}\\
        $p$ & {\color{toPat}$\stepsf$} & {\color{toPat} $p'$}\\
        \hspace{1em}$\of{n+1}{}{}$& &{\color{toStep}$\of{n}{}{}$}\\
        $d$ & {\color{toStep}$\steps$} & {\color{toStep}$d'$}
    \end{tabular}
    \caption{Combination of {\color{toPat}\cref{lemma:step_stepf}} and {\color{toStep}\cref{lemma:stepf_step}} assuming $e\terminates$}
    \label{fig:ofpreserve}
\end{figure}

The rest of the lemmas are highlighted in \cref{fig:ofpreserve}. We start with the givens $e\steps e'$, $\of{0}{e}{p}$, and $\of{n+1}{d}{p}$, then we acquire ${\color{toPat}p\stepsf p'}$ and ${\color{toPat}\of{0}{e'}{p'}}$ from {\color{toPat}\cref{lemma:step_stepf}}, then we conclude ${\color{toStep}d\steps d'}$ and ${\color{toStep}\of{n}{d'}{p'}}$ from {\color{toStep}\cref{lemma:stepf_step}}. These steps require the main bulk of the formalization proof effort as they will involve the specific cases of interest of calling the function in question.

\begin{lemma}[ {\color{toPat}$\steps$ to $\stepsf$} ]\label{lemma:step_stepf} If $e\terminates$ and $\of{0}{e}{p}$ and $e\steps e'$ then there exists $p'$ such that $p\stepsf p'$ and $\of{0}{e'}{p'}$.
\begin{proof}
By induction on $e\steps e'$ and inversion on $\of{n}{e}{p}$. Most of the cases are trivial as $\stepsf$ contains all the same rules as $\steps$, and we can utilize \cref{lemma:val_valf} and the induction hypothesis to convert between them.

For the case of $e=\tmapp{F_{\omega}}{v}$ and $p=\tmapp{\fhole}{v}$, we can simply apply the $\fhole\stepsf$ rule and continue the proof from there.

For the case of $e=\tmapp{F_0}{v}$, we utilize \cref{lemma:inf_loop} to show this case contradicts $e\terminates$.

For the case of $e=\tmapp{F_{i+1}}{v}$ and $p=\tmapp{\fhole}{v}$, we can simply apply the $\fhole\stepsf$ rule and continue the proof from there, as we are guaranteed at least one unrolling.
\end{proof}
\end{lemma}

Note that \cref{lemma:step_stepf} relies on the assumption of $\of{0}{e}{p}$ and provides a result of $\of{0}{e'}{p'}$ in the conclusion.
We may strengthen the conclusion to $\of{n}{e'}{p'}$ for any natural number $n$ desired, as long as we strengthen the assumption to be that of $\of{n+1}{e}{p}$, thus removing the need for the $e\terminates$ judgment.
However, this generalization is not sufficient for the backward direction of compactness $\left(\exists n.\ \subst{F_n}{x}{e}\terminates\right)\ implies\ \subst{F_\omega}{x}{e}\terminates$ since it fails several edge cases involving the possibility where $n=0$ and the function is never called during execution.
It is easier to consolidate these cases into this single lemma as stated since the weaker conclusion of $\of{0}{e'}{p'}$ is all that is required in this overall proof strategy.

\begin{corollary}[$\steps^n$ to $\stepsf^n$]\label{lemma:step_stepf_multi} If $\of{0}{e}{p}$ and $e\steps^n e'$ and $e'\ \val$ then there exists $p'$ such that $p\stepsf^n p'$ and $\of{0}{e'}{p'}$.
\end{corollary}

Having done all the effort to convert into $\stepsf$, we now reap the benefits of having these empty holes in the pattern which we can fill in with whatever unrolling is desired. The preservation of the pattern structure for this arbitrary filling is expressed in the following lemma.

\begin{lemma}[ {\color{toStep} $\stepsf$ to $\steps$} ]\label{lemma:stepf_step} If $\of{n+1}{d}{p}$ and $p\stepsf p'$ then there exists $d'$ such that $d\steps d'$ and $\of{n}{d'}{p'}$.
\begin{proof}
By induction on $p\stepsf p'$ and inversion on $\of{n+1}{d}{p}$. Once again, most of the cases are trivial as we can utilize \cref{lemma:val_valf} and the induction hypothesis to convert between $\stepsf$ and $\steps$. We utilize \cref{lemma:dec_of} to decrease the minimum required depth of unrollings by one whenever applicable.

For the case of $d=\tmapp{F_{\omega}}{v}$ and $p=\tmapp{\fhole}{v}$, we can simply apply the $\Tfun{}{}\steps$ rule and continue the proof from there.

The case of $d=\tmapp{F_i}{v}$ and $p=\tmapp{\fhole}{v}$ for $i<n+1$ contradicts $\of{n+1}{d}{p}$ as the depth of unrolling is less than the stated minimum of $n+1$.

For the case of $d=\tmapp{F_i}{v}$ and $p=\tmapp{\fhole}{v}$ for $i\geq n+1$, we can simply apply the $\Tfun{}{}\steps$ rule and continue the proof from there, as we are guaranteed at least one unrolling. We still have the desired minimum recursive depth since $i-1\geq n$.
\end{proof}
\end{lemma}

\begin{corollary}[$\stepsf^n$ to $\steps^n$]\label{lemma:stepf_step_multi} If $\of{n}{d}{p}$ and $p\stepsf^n p'$ then there exists $d'$ such that $d\steps^n d'$ and $\of{0}{d'}{p'}$.
\end{corollary}

Note that this $\of{n}{d}{p}$ assumption for \cref{lemma:stepf_step_multi} is dependent on the total number $n$ of steps taken. For most cases, it is possible to decrease the strength of the assumption to be lower than $n$, but proving such a lemma would require the excessive bookkeeping we are trying to avoid and is unnecessary to prove our desired version of compactness.

We are now finally able to prove a generalized notion of compactness that follows very naturally from the combination of all the lemmas we have in place. This lemma is strictly stronger than our desired definition of compactness in \cref{lemma:compactness} as we know by definition that $\of{i}{\subst{F_i}{\fhole}{e}}{e}$.

\begin{lemma}[Generalized Compactness]\label{lemma:generalized} If $e\terminates$ in $n$ steps, then for all $p$ and $d$ such that $\of{0}{e}{p}$ and $\of{n}{d}{p}$, we know $d\terminates$ in $n$ steps.
\begin{proof}
    First, decompose $e\downarrow$ into $e\steps^n e'$ and $e'\ \val$.
    Then, apply \cref{lemma:step_stepf_multi} to get $p\stepsf^n p'$ where $\of{0}{e'}{p'}$.
    From here, we can apply \cref{lemma:val_valf} to get that $p'\ \valf$.
    Next, we apply \cref{lemma:stepf_step_multi} to get $d\steps^n d'$ and $\of{0}{d'}{p'}$.
    We finish by showing $d'\ \val$ via \cref{lemma:val_valf}.
\end{proof}
\end{lemma}

\section{Coq Formalization with Continuations}

So far in this paper, we have discussed the generic overall proof strategy for proving compactness, but have not covered the specifics of how it can be formally verified in the setting of explicit control flow.
In the remainder of the paper, we go into deeper detail about our specific application of this proof strategy and how it can be formally verified in Coq.
All proofs were verified in version 8.15.2 of Coq.
The formulation is linked here: \href{https://github.com/pi314mm/compactness}{https://github.com/pi314mm/compactness}.

Reasoning about programs with explicit control effects adds an extra layer of difficulty to the formalization in every aspect.
The methodology of proving compactness expressed in this paper extends to a language with letcc/throw, and we made several important design choices in the definition of the language and lemmas to prove things cleanly.
We shall closely examine how several lemmas throughout the formalization change slightly to handle this extended case.

We will find that most of the burden of the Coq formalization is associated with shortcomings in handling substitution goals as we see general patterns of similar proofs of substitution appearing over and over again that should be obvious on paper but not in Coq.
This is made even more apparent in the realm of continuations since we frequently must reason about substitutions into closed programs and closed types: it is trivial to conclude that the substitution does not affect the closed term, but in practice, we must carry out the full inductive proof.

\subsection{Language Specification}

The first design decision we had to make was choosing between different representations of stack frames and evaluation contexts.
Harper \cite{harper2016practical} presents one plausible formulation that is very explicit in how values are returned to stack frames by making use of two different judgments:
$k \triangleright e$ to represent that term $e$ has computations to be done, and $k\triangleleft v$ to represent that term $v$ is a value and is returning up the stack.
This formulation has the benefit that the continuation component is kept separate from the term component, so the next step to take in execution is always immediately obvious, making dynamic stepping rules require no premises.
However, we opted against this formulation due to the excess machinery involved in making multiple judgments, and also due to the extra small steps needed to make phasing between the two judgments explicitly.

We instead choose to utilize evaluation contexts $E[e]$ for continuation $E$ and term $e$ with a slight twist on the definition of the dynamics.
Normally with evaluation contexts, the inductive cases of the stepping relation are made implicit, utilizing a two-part relation $E[e]\steps e'$ where $e$ is a redex.
Instead, we utilize a three-part relation $E; e\steps e'$ and explicitly define the inductive cases of the stepping relation by transferring stack frames from $e$ to $E$.
This gives us the best of both approaches where we have better control over the inductive cases through stack frames, but we don't have to deal with the secondary judgment of returning a value to a continuation.

\begin{figure}
    \centering
    \begin{align*}
        \tau &:= 0\mid 1\mid \tau \to \tau \mid \tau\times\tau\mid \forall x.\tau \mid \exists x.\tau \mid \Tcont{\tau}\\
        e &:=\tmlet{e}{x}{e}\mid \tmabort{\tau}{e} \mid ()\\
        &\quad\mid \tmfun{f}{x}{\tau}{\tau}{e} \mid e\ e \mid \tmpair{e}{e}\mid \tmprojl{e}\mid \tmprojr{e}\\
        &\quad\mid \tmLam{x}{e} \mid \tmApp{e}{\tau} \mid \tmpack{\tau}{e}{x}{\tau} \\
        &\quad \mid \tmopen{e}{x}{y}{e} \mid \tmletcc{\tau}{x}{e}\\
        &\quad \mid \tmthrow{e}{e} \mid \tmcont{e}\\
        E &:= \hole
    \end{align*}
    \caption{Full grammar of language}
    \label{fig:grammar}
\end{figure}

In addition to recursive function and explicit control flow, we add various other features into our language including polymorphic types.
The complete grammar of our language is shown in \cref{fig:grammar}.
This syntax was defined as one inductive datatype within Coq which meant we could reuse the syntax from term to define our evaluation context stack frames, hence only needing the one extra base case, and we only need to implement substitution once to cover substitutions into all three constructs.
The syntax can be found in the SyntaX.v file of the Coq supplement.

Variables in the language are implemented using De Bruijn indices.
We utilized the explicit substitution framework implemented by Crary \cite{crary2019fully} with some modifications for ease of use.
This framework gives us various tools for substitution in a generalized setting which we use to simplify several substitution goals throughout our formalization.
The files related to this can be found in the EasySubst folder of the Coq supplement.

\begin{figure}
    \centering
    \begin{mathpar}
    \inferrule{
        \static{\Gamma, x : \Tcont{\tau}}{e}{\tau}
    }{
        \static{\Gamma}{\tmletcc{\tau}{x}{e}}{\tau}
    }\and
    \inferrule{
        \static{\Gamma}{e_1}{\tau}\\
        \static{\Gamma}{e_2}{\tmcont{\tau}}
    }{
        \static{\Gamma}{\tmthrow{e_1}{e_2}}{0}
    }\and
    \inferrule{
        \ESJ{E}{\tau}{1}
    }{
        \static{\Gamma}{\tmcont{E}}{\Tcont{\tau}}
    }\and
    \inferrule{
    }{
        \tmcont{E}\ \val
    }\and
    \inferrule{
    }{
        E;\tmletcc{\tau}{x}{e} \steps E[\subst{\tmcont{E}}{x}{e}]
    }\and
    \inferrule{
        E[\tmthrow{\hole}{e_2}]; e_1 \steps e'
    }{
        E;\tmthrow{e_1}{e_2} \steps e'
    }\and
    \inferrule{
        e_1\ \val\and
        E[\tmthrow{e_1}{\hole}]; e_2 \steps e'
    }{
        E;\tmthrow{e_1}{e_2} \steps e'
    }\and
    \inferrule{
        e\ \val
    }{
        E;\tmthrow{e}{\tmcont{E'}} \steps E'[e]
    }
    \end{mathpar}
    \caption{Selected statics and dynamics}
    \label{fig:letccthrowsemantics}
\end{figure}

\Cref{fig:letccthrowsemantics} illustrates the static and dynamic typing judgments associated with explicit control flow.
For the static typing judgments, the important things to note are that $\tmthrow{e_1}{e_2}$ is of type $0$ and that $\tmcont{E}$ requires that $E$ be an evaluation context converting from $\tau$ in the empty context to $1$ in the empty context.
The static and dynamic typing judgments can be found in the Rules.v file of the Coq supplement.

We opted to restrict our $\tmthrow{e_1}{e_2}$ to the type $0$ since we have $\tmabort{\tau}{e}$ in our language capable of changing the type $0$ into whatever we want.
Marking $\tmthrow{e_1}{e_2}$ as the type $0$ helps us identify at a type level that the computation will never return a value to the current evaluation context, allowing for certain lemmas outside the scope of compactness to hold vacuously true.

The reason for the restriction on the evaluation context having a return type of $1$ is how it affects the safety proof of the language.
We restrict our definition of closed, whole programs to being closed terms of the answer type $1$, meaning that all evaluation contexts produced during evaluation have the return type of $1$.
Enforcing this via the statics eliminates the need to maintain this as an invariant in the proof of type safety, making it a lot easier to formally verify within Coq.

The small-step semantics for $\tmletcc{\tau}{x}{e}$ and the last case of $\tmthrow{e}{\tmcont{E'}}$ are standard rules when dealing with continuations as the three-part relation version doesn't look much different from the two-part relation version.
The two other inductive rules for $\tmthrow{e_1}{e_2}$ demonstrate how stack frames from the computation are transferred over to the evaluation context explicitly, allowing us greater control over these inductive cases when we induct over the stepping relation.

The next step we take is to prove several theorems about substitutions in the typing judgments.
The generalized substitution framework provides some tools for reasoning about substitutions, but we need to adapt it to the setting of explicit control flow, specifically for the typing judgment for $\tmcont{E}$ since it requires reasoning about substitutions into closed terms.
This is an excellent example of a lemma that seems trivial on paper but requires nontrivial effort when expressed in Coq.
With some effort, we are able to formally verify the following necessary lemma about substituting in a closed term to finish off the overall substitution lemmas.

\begin{lemma}[Empty Context Substitution]\label{lemma:sub_eq}\quad
    \begin{enumerate}
        \item $\cdot\vdash \tau\ implies\ \sigma(\tau)=\tau$
        \item $\static{\cdot}{e}{\tau} \ implies\  \sigma(e)=e$
        \item $\ESJ{E}{\tau}{1} \ implies\  \sigma(E)=E$
    \end{enumerate}
\end{lemma}

These substitution-related lemmas can be found in the Substitution.v file of the Coq supplement.

With everything in place, we are finally able to tackle the type safety theorem.
However, in the setting of explicit control flow, the theorem looks slightly different than one would normally expect.

\begin{theorem}[Progress]If $\static{\cdot}{e}{\tau}$ then either $e\ \val$ or $\forall \ESJ{E}{\tau}{1}.\ \exists e'.\ E;e\steps e'$\end{theorem}
\begin{theorem}[Preservation]If $\static{\cdot}{e}{\tau}$ and $\ESJ{E}{\tau}{1}$ and $E;e\steps e'$ then $\static{\cdot}{e'}{1}$\end{theorem}

For progress, instead of just stating there is a next step, we must consider all possible enclosing evaluation contexts that could enclose this term and show that there exists a next step for each one of them.
This strengthens the inductive hypothesis enough to carry out the proof.

For preservation, we are given the additional assumption about the evaluation context having a valid type on the left, and we conclude that the right side is a closed program of type $1$ in accordance with the three-part relation.

Combining these two theorems together gives us the type safety theorem, fulfilling the requirement expressed by \cref{lemma:safety}.
These theorems can be found in the Saftey.v file of the Coq supplement.

The other requirement of determinism, \cref{lemma:determinism}, is fairly straightforward to prove for this language.
This is proved in the Rules.v file of the Coq supplement.
This determinism lemma is only utilized for \cref{lemma:inf_loop}, which can be found in the Kleene.v file of the Coq supplement.

\subsection{Formalizing Pattern Language}

The remainder of the proofs covered in this paper can all be found within the Compactness.v file of the Coq supplement.

The biggest design decision involved in creating the pattern language associated with our chosen language is how to implement the open global variable $\fhole$.
Do we want to handle substitution into $\fhole$ in the same way as local variables but with a global scope or express it as a different syntactic construct?

We originally tried to express $\fhole$ as the $0$th variable per our De Bruijn indices interpretation of variables.
The benefit of this is that we did not need to implement any new instrumentation to handle substitution into the $\fhole$ variable as we could just utilize the same framework we use for local variables.
The disadvantage is that we need to consider the index of the variable changing when going underneath binders within the term.
As such, the $\mathbf{of}$ judgment needs to carry an extra counter indicating what variable index is associated with $\fhole$, and this counter changes within the inductive definition of the $\mathbf{of}$ judgment whenever entering into a bound scope.
The substitution lemmas associated with this setup proved to be rather bulky as we essentially had to reprove the substitution lemma in a way that singled out the $0$th variable, and it unnecessarily complicated the definition of the $\mathbf{of}$ judgment, although this method does work well with good enough support for variable bindings.

We then scrapped that idea and instead defined the ``pat'' constant in our language to represent the variable $\fhole$.
We still had to prove substitution-related theorems for this setting, but it was a lot simpler as we only had to consider a single isolated variable instead of needing to reindex De Bruijn indices.

\begin{figure}
    \begin{minted}{Coq}
Inductive Of (A B f : term) (n : nat) :
    term -> term -> Prop :=
...
| Of_tm_fun : forall e1 e2 e3 p1 p2 p3,
  Of A B f n e1 p1 ->
  Of A B f n e2 p2 ->
  Of A B f n e3 p3 ->
  Of A B f n (tm_fun e1 e2 e3) (tm_fun p1 p2 p3)
...
| Of_pat_inf :
    Of A B f n (tm_fun A B f) pat
| Of_pat_fin : forall i, i >= n ->
    Of A B f n (unroll i A B f) pat
    \end{minted}
    \caption{Select cases of the $\mathbf{of}$ judgment}
    \label{fig:of}
\end{figure}

Beyond the necessary substitution lemmas, we also needed to implement a lot of intermediate lemmas dealing with $\mathbf{of}^n$ including the following lemmas.

\begin{lemma}[Compose $\mathbf{of}^n$]\label{lemma:compose}If $\of{n}{E}{P}$ and $\of{n}{e}{p}$ then $\of{n}{E[e]}{P[p]}$.\end{lemma}

\begin{lemma}[Reflexivity for $\mathbf{of}$]\label{lemma:reflex_of}\quad
\begin{itemize}
    \item $\Gamma\vdash \tau\ \mathrm{implies}\ \of{n}{\tau}{\tau}$
    \item $\static{\Gamma}{e}{\tau}\ \mathrm{implies}\ \of{n}{e}{e}$
    \item $\ESJ{E}{\tau}{1}\ \mathrm{implies}\ \of{n}{E}{E}$
\end{itemize}
\end{lemma}

\subsection{Compactness}

In order to prove compactness with the extended three-part relation for continuations, several intermediate steps need to be extended to account for open evaluation contexts with variable $\fhole$.

We need to make the following modifications to \cref{lemma:step_stepf,lemma:stepf_step} by generalizing over their evaluation context.

\begin{lemma}[ {\color{toPat}$\steps$ to $\stepsf$} with Continuations]\label{lemma:step_stepf_cont} If $E[e]\terminates$ and $\of{0}{E}{P}$ and $\of{0}{e}{p}$ and $E; e\steps e'$ then there exists $p'$ such that $P; p\stepsf p'$ and $\of{0}{e'}{p'}$.
\end{lemma}

\begin{lemma}[ {\color{toStep} $\stepsf$ to $\steps$} with Continuations]\label{lemma:stepf_step_cont} If $\of{n+1}{D}{P}$ and $\of{n+1}{d}{p}$ and $P; p\stepsf p'$ then there exists $d'$ such that $D; d\steps d'$ and $\of{n}{d'}{p'}$.
\end{lemma}

The overall proof strategy for proving these lemmas remains the same, we simply need to manage the overhead of evaluation contexts using \cref{lemma:compose} whenever needed.

\Cref{lemma:val_valf} for relating values does not change at all.
The multi-step \cref{lemma:step_stepf_multi,lemma:stepf_step_multi} and even the generalized compactness \cref{lemma:generalized} also do not change since the multi-step relation for continuations is defined by chaining single steps of the form $\hole; e\steps e'$ for the empty evaluation context $\hole$, and we can show that $\of{n}{\hole}{\hole}$ by definition.
The final desired lemma of compactness does not change significantly: we simply add evaluation contexts to the definition.
This is where \cref{lemma:reflex_of} comes into play by granting us that $\of{0}{E}{E}$.

\begin{lemma}[Compactness with Continuations]\label{lemma:compactness_cont}$E[\subst{F_\omega}{x}{e}]\terminates\ \mathrm{iff}\ \exists n.\ E[\subst{F_n}{x}{e}]\terminates$.\end{lemma}

As we can see, even when dealing with this modified setting with control flow effects, the overall proof strategy remains largely the same.
Adding more features into the language even beyond continuations does not present a significant impact on the compactness lemma, as the general pattern of execution remains the same upon replacing instances of the recursive function.

\section{Necessity of Pattern Stepping}

At first, the pattern-stepping formulation might seem to be a convenient hack for formally verifying compactness, and it seems to come out of nowhere.
This begs the question of why we are using pattern-stepping in the first place, especially considering other approaches to proving compactness have worked in the past without it.

To answer this, consider what it would take to formally verify the compactness lemma with as straightforward induction as possible.
Two things change within the execution of the program concerning the recursive function in question: the depth of unrolling and the locations of the function, so our inductive hypothesis would need to account for those two things.

The depth of the unrollings can either all be accounted for, or we can simply account for the minimum.
Choosing the latter gives us something along the lines of the relation $\mathbf{of}^{n}$ which is very easy to formulate into an inductive hypothesis.

Now, specifying the locations of the recursive functions, especially within a proof assistant, adds an additional layer of difficulty.
It is insufficient to make a syntactic distinction: within the statement of the compactness lemma, the term $\subst{F_\omega}{x}{e}$ imposes no restrictions on the definition of $e$, so there could be syntactically identical copies of $F_\omega$ or its unrollings within the term.
As such, it becomes necessary to mark out the places within $e$ which are relevant to the compactness lemma, leading us to the pattern language.

Another way of seeing the pattern term is as a multi-context $C\{x_1,x_2,x_3,\cdots\}$ which is a part of the Pitts formulation of compactness \cite{pitts1997operationally}.
They both serve the same purpose in keeping track of the locations of the recursive function within the term, however, the pattern term is much easier to work with within a proof assistant than is the multi-context, and an attempt to formalize a multi-context within a proof assistant would result in something rather similar to a pattern term.

Taking a look at \cref{fig:ofpreserve}, it is possible to skip the pattern language formulation by using a three-part relation that might look like $(e,p,d)\in\mathbf{of}^n$.
This utilizes the pattern $p$ to mark the places in $e$ and $d$ that are relevant to the recursive function in question, but it compresses the two lemmas of converting to the pattern language and back into a singular larger lemma.
However, the pattern language formulation gives meaning to the intermediate step in this lemma and chunks up the lemma into components that are easier to formally verify.

Our criteria for choosing the particular proof strategy of using a pattern language comes with formal verification in mind.
This formulation provides a very straightforward approach to compactness and most of the proof overhead incurred is easy to handle within a proof assistant.

\section{Generality of Pattern Stepping}

While we only presented a single (albeit complex) setting in which this general proof strategy can be used to prove compactness, we still argue that it is general enough for any programming language one can define through operational semantics.
The reason for this is that the proof technique is driven almost purely by the dynamics of the language, making almost no assumptions about the type theory of the language, or any of the technical components and constructors of the language, as all we care about are reduction semantics.
As such, this proof technique can be applied to any compactness lemma with almost no modification by simply handling the extra trivial compatibility cases.

We cannot formally prove that this proof strategy works for every language, as we cannot formalize the set of all languages, but we will express a few interesting extensions to this approach to argue for the generality of the proof technique.

\subsection{Dependent and Polymorphic Types}

This proof strategy does cleanly extend to the dependently typed setting and polymorphic typed setting without any modification.
The reason for this is that the dynamics of the term are not dependent on the type, as the type is only used for static checking before execution.

Our Coq implementation demonstrates this method works for polymorphic types.

\subsection{Mutual Recursive Functions}

Bekić's theorem \cite{bekic2005definable} allows us to extend the analysis of a single recursive function to that of mutual recursive functions.
However, we find that our proof framework still applies to the mutual recursion setting with a small modification.

Specifically, whereas before we only had one recursive function $F_\fhole$ and one pattern variable $\fhole$, we now have $n$ recursive functions $F_{\fhole,n}$ and $n$ pattern variables $\fhole_n$ to keep track of.
The minimum depth of unrolling can be shared among all of these functions, so the interface of the $\mathbf{of}^n$ judgment remains unchanged in all proofs, but we must add these extra cases into the definition of the $\mathbf{of}^n$ judgment.

This adds a small amount of overhead to the substitution lemmas for $\mathbf{of}^n$ and adds some base cases to \cref{lemma:step_stepf,lemma:stepf_step}, but these changes are straightforward to make.

\subsection{Syntax-Aware Code}

The one extreme case where our proof strategy does not work is the case where the programming language can read and case on its syntax during execution.
In this setting, program equivalence is trivialized to syntactic equivalence, making it generally uninteresting to reason about.

The compactness lemma is not true in such a setting, so we can still argue that our proof strategy works whenever the compactness lemma is true.

\section{Related Work}

The use of the compactness lemma has early roots in the literature as a means of converting domain-theoretic verification techniques of denotational semantics into the setting of operational semantics.
Through personal correspondence, we learn that the lemma was first coined as the ``compactness'' property by Harper as early as 1995.
Mason, Smith, and Talcott were the first to utilize the compactness lemma for operational semantics and proved the lemma by focusing on the specific one-step reduction relevant to the function approximation \cite{mason1996operational}.
The name ``compactness'' was adopted by Pitts in his development of a proof strategy for it using cofinal sets \cite{pitts1997operationally} in the realm of natural semantics. Soon later, this strategy was adapted to an operational semantics setting by Birkedal and Harper \cite{birkedal1999relational}.

The Pitts strategy involves formalizing a program context $C\{F_{n_1},F_{n_2},\cdots, F_{n_k}\}$ which takes in as input a vector that consists of various different levels of unrollings of the function in question.
The strategy relies on bookkeeping all these unrollings and future expansions to the unrolling vector during the overall computation process.
The important thing to note is that Pitts suggests that the best way to facilitate compactness is to reason about the whole program during execution, and does this via the program context.
Our formalization also relies on reasoning about the whole program but does so by maintaining invariants over the whole program.

Around the same time, Crary devised a different strategy to prove compactness utilizing applicative approximation \cite{crary1998type}.
Given closed computations $e_1$ and $e_2$ of the same type, the applicative approximation $e_1 \preceq e_2$ means that if $e_1$ reduces to a value $v_1$, then $e_2$ must also reduce to a value $v_2$ where $v_1 \preceq_{val} v_2$ for some language-specific definition of $\preceq_{val}$.
With this, it is easy to see that $\bot$ approximates everything, so by extension, lower depths of unrollings approximate higher depths of unrollings.

The advantage of applicative approximation is that we no longer bookkeep all of the recursive unrollings during computation and we can instead simply keep track of the minimum depth of unrolling since all other unrollings are approximated by it.
The basis for what makes this a good approach is found within the idea of simulating multi-step reductions with a different but related relation that is agnostic to the exact depth of the unrollings.

Unfortunately, applicative approximation only works for negative connectives and does not generalize to positive connectives.
To handle positive connectives, Crary extends the approach utilizing contextual approximation \cite{crary2019fully}.
Unlike applicative approximation, it is not obvious that evaluation implies contextual approximation nor is it obvious that $\bot$ contextually approximates everything, so proving these theorems requires extra work.
Once these are proven for the specific language in question, the proof of compactness proceeds in the same way for both applicative and contextual approximations as they follow the basic principle of a relation that is agnostic to the exact depth of the unrollings and keep track of only the minimum such depth.

A large disadvantage of this proof strategy through contextual approximation is that it is not immediately obvious how it can be extended to the case of explicit control flow with continuations.
It is no longer the case that evaluation implies contextual approximation: we have the single-step evaluation of $\tmthrow{()}{\hole}\steps ()$, but the context $\tmlet{x}{\hole}{\bot}$ distinguishes between these two terms.
Our new proof strategy for compactness builds on the similar flavor of relating evaluation to a relation agnostic to unrollings, but unlike the previous methods, can cleanly extend to the setting of explicit control flow.

\subsection{Step-Indexed Logical Relations}

An alternative to compactness for FTLR is using step indexing to formalize the definition of logical relations.
The logical relation is then not only inductively defined on the type structure but also on an extra counter based on the occurrences of effectful operations (such as non-termination) in computation.
This counter strengthens the inductive hypothesis of FTLR, allowing us to ignore the issues presented by analyzing recursive functions.

The step-indexed model was first invented by Appel and McAllester \cite{appel2001indexed}, but was soon after adapted into the realm of logical relations for various settings by Ahmed and coworkers \cite{acar2008imperative,ahmed2006step,ahmed2009state,neis2009non}.

However, using step indexing has many drawbacks as it makes the logical relation less elegant to use after proving FTLR, as
instead of just using the type of the term, it forces us to keep track of an additional counter.
In this sense, step indexed logical relations are weaker than regular logical relations.
We argue that it is cleaner to separate the counter from the definition of the logical relation and instead only have it present within the proof of FTLR as is done via compactness.

It is reasonable to view our proof of compactness as a step-indexed bisimulation proof.
In this sense, we manage to separate the step-indexing component from the definition of the logical relation so that we only need to prove step-indexing once, and then benefit from this proof everywhere the logical relation is used.

There are additionally several variations of step-indexed logical relations that avoid the traditional presentation by instead incorporating a future modality \cite{dreyer2011logical}.
We will still categorize these as step-indexed logical relations because, while they greatly help make step-indexing more practical, they don't avoid the step-indexing issue within the definition of the logical relation as compactness allows.

\section{Conclusion}

As Pitts had claimed, the best way to deal with the compactness lemma is by analyzing the program as a whole during its execution.
The merit of this paper is not so much the ideas surrounding compactness as we draw upon many years of previously known insight, but the novelty of this paper comes from a rehashing of these ideas in a new way that makes compactness easy to formally verify in a proof assistant. 
In our setting when we are given the path of execution through small-step semantics, that is our whole computation, so maintaining the invariant of the $\mathbf{of}$ judgment over this whole term is what makes this proof work cleanly.

The ideas drawn from program capsules provide insight into separating the computable components of a program from the terminated value components of the program.
Adapting this concept to the realm of compactness gives us the pattern language described in this paper, which provides a clean formalization of the $\mathbf{of}$ invariant.
What we truly care about are the computable components of the program and the overall pattern of computation they represent, while abstracting over the specific depth of the unrolling, and the pattern language helps us describe precisely that.

When translating these proofs into a proof assistant like Coq, several implicit details (mostly pertaining to substitution) that could have been hand-waved in paper proofs become non-trivial, formally verified proofs.
Unlike some of the earliest work on compactness, this proof strategy was developed with formal verification in mind, and thus the overall proof strategy is clean in both the paper version of lemmas and the machinery utilized to formally verify them.

\bibliographystyle{plain}
\bibliography{ref}

\end{document}